\documentclass[10pt,journal,twocolumn,final]{IEEEtran}
\usepackage{multind}
\usepackage[T1]{fontenc}
\usepackage{ae,aecompl}
\usepackage[utf8]{inputenc}
\usepackage[english]{babel}
\usepackage{amsmath}
\usepackage{amsfonts}
\usepackage{amssymb}
\usepackage{mathtools}
\usepackage{amsthm}
\usepackage{times}
\usepackage{verbatim, amsbsy}
\usepackage{graphicx}
\usepackage{epsfig}
\usepackage[normalem]{ulem}
\usepackage{color}
\usepackage{fancyhdr}
\usepackage{lastpage}
\usepackage{listings}
\usepackage{transparent}
\usepackage{subfigure}
\usepackage{bm}
\usepackage[all]{xy}
\usepackage[normalem]{ulem}

\newfont{\bbb}{msbm10 scaled 500}

\newfont{\bb}{msbm10 scaled 1100}

\newcommand{\bv}{{\bf b}}

\newcommand{\hv}{{\bf h}}

\newcommand{\nv}{{\bf n}}

\newcommand{\qv}{{\bf q}}
\newcommand{\rv}{{\bf r}}
\newcommand{\sv}{{\bf s}}

\newcommand{\uv}{{\bf u}}

\newcommand{\xv}{{\bf x}}
\newcommand{\yv}{{\bf y}}

\newcommand{\Am}{{\bf A}}
\newcommand{\Bm}{{\bf B}}

\newcommand{\Em}{{\bf E}}
\newcommand{\Fm}{{\bf F}}

\newcommand{\Hm}{{\bf H}}
\newcommand{\Id}{{\bf I}}

\newcommand{\Lm}{{\bf L}}

\newcommand{\Qm}{{\bf Q}}

\newcommand{\Wm}{{\bf W}}

\newcommand{\Bc}{{\cal B}}
\newcommand{\Cc}{{\cal C}}

\newcommand{\Nc}{{\cal N}}

\newcommand{\Tc}{{\cal T}}

\newcommand{\phiv}{\hbox{\boldmath$\phi$}}

\newcommand{\Phim}{\hbox{\boldmath$\Phi$}}

\newcommand{\trace}{{\hbox{tr}}}

\newcommand{\herm}{^{\text{H}}}

\newcommand{\mx}{{\text{m}}}

\newcommand{\sx}{{\text{s}}}

\newcounter{lecnum}
\newcommand{\executeiffilenewer}[3]{%
\ifnum\pdfstrcmp{\pdffilemoddate{#1}}%
{\pdffilemoddate{#2}}>0%
{\immediate\write18{#3}}\fi%
}

\DeclareMathOperator{\rank}{rank}

\DeclareMathOperator{\dimV}{dim}
\DeclareMathOperator{\kerV}{ker}

\begin{document}

\title{Cognitive Orthogonal Precoder for Two-tiered Networks Deployment}

\author{Marco Maso, \IEEEmembership{Student Member, IEEE,
} Leonardo S. Cardoso, \IEEEmembership{Member, IEEE,} M\'erouane~Debbah,~\IEEEmembership{Senior~Member,~IEEE,} and Lorenzo Vangelista, \IEEEmembership{Senior Member, IEEE}%
\thanks{Manuscript received date: Apr 16, 2012; Manuscript revised date: Sep~5,~2012. This work was partially supported by the ERC Starting Grant 305123 MORE (Advanced Mathematical Tools for Complex Network Engineering).}%
\thanks{M. Maso and M. Debbah are with the Alcatel-Lucent Chair on Flexible Radio - SUP\'ELEC, Gif-sur-Yvette, France (e-mail: {marco.maso, merouane.debbah}@supelec.fr).}
\thanks{L. S. Cardoso is with INRIA, Université de Lyon, INSA-Lyon, CITI-INRIA, France (e-mail: leonardo.sampaio-cardoso@inria.fr).}
\thanks{L. Vangelista is with the Department of Information Engineering, University of Padova, Italy (e-mail: lorenzo.vangelista@unipd.it).}}

\maketitle
   
\begin{abstract}

In this work, the problem of cross-tier interference in a two-tiered (macro-cell and cognitive small-cells) network, under the complete spectrum sharing paradigm, is studied. A new orthogonal precoder transmit scheme for the small base stations, called multi-user Vandermonde-subspace frequency division multiplexing (MU-VFDM), is proposed. MU-VFDM allows several cognitive small base stations to coexist with legacy \linebreak macro-cell receivers, by nulling the small- to macro-cell cross-tier interference, without any cooperation between the two tiers. This cleverly designed cascaded precoder structure, not only cancels the cross-tier interference, but avoids the co-tier interference for the \linebreak small-cell network. The achievable sum-rate of the small-cell network, satisfying the interference cancelation requirements, is evaluated for perfect and imperfect channel state information at the transmitter. Simulation results for the cascaded \linebreak MU-VFDM precoder show a comparable performance to that of \linebreak state-of-the-art dirty paper coding technique, for the case of a dense cellular layout. Finally, a comparison between \linebreak MU-VFDM and a standard complete spectrum separation strategy is proposed. Promising gains in terms of achievable \linebreak sum-rate are shown for the two-tiered network w.r.t. the traditional bandwidth management approach.

\end{abstract}
\section{Introduction} \label{sec:intro}

Recent academic and industry trends point towards a paradigm shift in wireless communications: the adoption of a two-tiered network structure (e.g., \cite{art:chandrasekhar08}). Two-tiered networks aim at breaking away from the traditional cellular layout to provide the expected capacity increase for future wireless services. As the name implies, its main difference from the traditional cellular paradigm, is the deployment of a second tier of densely populated and self-organizing small base stations (SBS)~\cite{art:hoydis11}. In spite of all these qualities, the addition of an SBS layer requires coexistence with the existing macro base station (MBS) infrastructure. 

Traditionally, coexistence in two-tiered networks is accomplished by means of three different approaches \cite{conf:andrews10}. In \textit{complete separation}, the MBSs and SBSs operate on disjoint bands, which avoids the mutual interference between the tiers, \linebreak i.e., \textit{cross-tier interference}, but decreases the spectral efficiency. To enhance spectral efficiency, the two tiers can share part of the total available band, under the \textit{partial sharing} paradigm. In order to work, solutions for cross-tier interference control need to be adopted for the shared band. For a maximal spectral re-use, the most attractive solution is the \textit{complete sharing}, where the MBSs and SBSs share all the band. Despite its notable features, this approach leads to an unbearable cross-tier interference, thus requiring interference management techniques for the best coexistence of the two network tiers.  

Out of the many techniques that exist for coexistence, a popular one named interference alignment (IA) \cite{art:cadambe2008i}, copes with cross-tier interference by isolating the received and interference signal subspaces. This requires a smart coordination of the devices in the network and special decoding at the receiver to realize the alignment. Interestingly, IA-based solutions for different channel state information (CSI) assumptions have been proposed, requiring the existence of exploitable degrees of freedom in the spatial \cite{conf:ghasemi10}, frequency \cite{conf:da11} or time \cite{conf:zhou10d} domain. Coordinated beamforming \cite{art:dahrouj10} based solutions require CSI at the transmitter only, with the advantage of no special decoding at the receiver. On the other hand, the power normalization needed at the transmitter to fulfill the transmit power constraints may result in performance penalties, depending on the condition number of the resulting channel matrices, as the number of involved MBSs/SBSs grows. Alternatively, in the absence of cooperation between the tiers, interference can be managed through dynamic spectrum access (DSA) \cite{art:zhao07}. DSA strategies such as spectrum shaping \cite{art:zhang10a} and cooperative frequency reuse \cite{art:akoum11} can be adopted at the SBSs, depending on the spectrum management approach adopted by the MBS.

Proposed for a similar problem, cognitive radios (CR) \cite{art:haykin2005} aim at fostering spectrum re-use by protecting a \emph{primary} (legacy) system from the interference generated by a \emph{secondary} (opportunistic) one. By labeling the MBSs as the primary system (first tier) and the SBSs as the secondary system (second tier), CR networks can be also seen as a particular case of two-tiered networks under the complete sharing approach, in which interference protection is inherently asymmetric (from the second to the first tier only). An application of IA to CR networks has been proposed in contributions such as \cite{conf:zhou11d}, where an SBS is aware of the power allocation performed in a multi-antenna MBS and aligns the cross-tier interference towards the macro-cell receivers. CR beamforming approaches have also been proposed in \cite{art:gastpar07, art:zhang08}, where multiple spatial dimensions at a cognitive transmitter are used to mitigate the cross-tier interference. This is accomplished by satisfying a signal to interference plus noise ratio (SINR) constraint at the primary receivers, while at the same time serving a reasonable rate to one or more secondary receivers.

In this contribution, we specifically target a two-tiered system comprised of a long term evolution (LTE) \cite{rpt:3gpp25.814} orthogonal frequency division multiple access (OFDMA) MBS and several cognitive SBSs operating under the complete sharing approach. The SBS system is modeled as a coordinated network multiple input multiple output (MIMO) system with infinite backhaul capacity. We propose a novel DSA CR technique for this scenario, called multi-user \linebreak Vandermonde-subspace frequency division multiplexing \linebreak (MU-VFDM). MU-VFDM consists of a cascaded linear precoder made up by an inner component designed to cancel the cross-tier interference from the SBSs to the first tier, and an outer component to avoid the multi-user interference in the second tier, i.e., \textit{co-tier interference}. We show that, not only OFDMA, but any block transmission scheme that deals with multipath interference, provides resources that can be exploited by MU-VFDM to cancel the cross-tier interference. Under this assumption, the sole requirement is perfect CSI at the transmitter (CSIT), used to derive the precoder. This contrasts with the aforementioned \linebreak state-of-the-art techniques, that either require available time, space or frequency resources, or cooperation between the tiers to be performed. Sum-rate enhancements are shown to be achievable w.r.t. to the legacy complete separation approach, for a large range of signal to noise ratio (SNR) values, regardless of the number of SBSs. Additionally, the impact of imperfect CSI at the transmitter is evaluated, providing important design insights.

This paper is organized as follows. In Sec. \ref{sec:model}, we introduce the general MBS/SBS model assumed throughout this paper. Then, we derive the precoders and briefly discuss their performance in Sec. \ref{sec:prec_des}. In Sec. \ref{sec:numerical}, we present some numerical results for our MBS/SBSs study case. Finally, conclusions and future research directions are discussed in Sec. \ref{sec:conclusion}.

In this work, we adopt the mathematical notation as described in the following. A lower case italic symbol (e.g. \textit{b}) denotes a scalar value, a lower case bold symbol (e.g. $\bv$) denotes a vector, an upper case bold symbol (e.g. $\Bm$) denotes a matrix. $[\Bm]_{m,n}$ denotes a matrix element at the $m^{\text{th}}$ row and the $n^{\text{th}}$ column. An $\Id_{N}$ denotes the identity matrix of size $N$. The transpose conjugate operator on a matrix is denoted by the $\text{H}$ superscript (e.g. $\Bm^{\text{H}}$), the transpose operator is denoted by the $\text{T}$ (e.g. $\Bm^{\text{T}}$), the Moore-Penrose pseudoinverse matrix is denoted by $\dagger$ (e.g. $\Bm^{\dagger}$), $\kerV{(\Bm)}$ denotes the kernel of the matrix $\Bm$ and $\trace(\Bm)$ its trace. The operator $\Am \otimes \Bm$ is used to represent the Kronecker product. The special matrix $\textbf{0}_{N,M}$ denotes the zero matrix of dimension $N \times M$. All vectors are columns, unless otherwise stated.

\section{System Model} \label{sec:model}

\begin{figure}[!h]
\centering
\includegraphics[width=0.92\columnwidth]{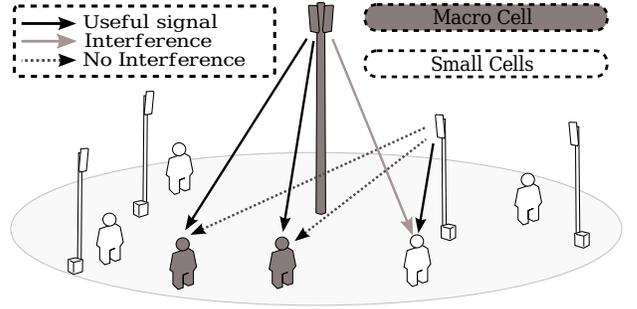}
	\caption{MU-VFDM downlink model, two-tiered network.}
	\label{fig:scenario}
\end{figure}

Consider the downlink scenario in Fig.~\ref{fig:scenario}, where all communications are assumed to be in time division duplex (TDD) mode. 
An MBS and $K$ cognitive SBSs, transmitting over the same frequency band, are deployed in a given area. The MBS serves $M$ single-antenna macro-cell user equipments (MUEs). The SBSs are considered to cooperate, yielding a full network MIMO transmission system model \cite{art:gesbert10}. For simplicity, and without loss of generality, we assume that each SBS serves one single-antenna small-cell user equipment (SUE). Concerning the notation, subscript ``$\text{m}$'' refers to the MBS, while ``$\text{s}$'' refers to the SBSs, i.e., $\hv_{\text{sm}}^{(i,j)}$ (or $\Hm_{\text{sm}}^{(i,j)}$) represents a link from SBS $i$ to MUE $j$. Conversely, $\sv^{[i]}$ (or $\Hm_{\text{sm}}^{([i],j)}$) denotes a vector/matrix related to the transmission from any SBS \emph{except} $i$. All channel vectors $\hv \in{\Cc}\Nc(0,\Id_{L+1}/(L+1))$, irrespective of the tier, transmitter and receiver, represent the impulse response of independent and identically distributed (i.i.d.) frequency-selective Rayleigh fading channels composed of $L+1$ paths.

The MBS adopts an $M$-user OFDMA based transmission of block size $N+L$ and a cyclic prefix of size $L$. For simplicity, a uniform resource allocation of $N/M$ subcarriers per MUE is adopted, $\Nc_{j}$ being the set of subcarrier indices assigned to the $j^{th}$ MUE with $\displaystyle \bigcup_{j=1}^{M}\Nc_{j} = \{1,\hdots,N\}$ and $\displaystyle \bigcap_{j=1}^{M}\Nc_{j} = \emptyset$. As a consequence, each MUE selects its own set of subcarriers by means of an $N \times N$ mask receiver filter $\Bm_j$, such that $\trace(\Bm_j)=N/M$ and $\displaystyle\sum\limits_{j=0}^M \Bm_j = \Id_N$, with $[\Bm_j]_{(n, n)} = 1$ when the subcarrier $n$ is allocated to the $j^{th}$ MUE and zero otherwise. Let $\Fm \in \mathbb{C}^{N\times N}$ be a unitary discrete Fourier \linebreak transform (DFT) matrix with $[\Fm]_{(k+1, l+1)}=\frac{1}{\sqrt{N}}e^{-i2\pi \frac{kl}{N}}$ for $k,l = \{ 0,\dots,N-1\}$ and $\Am$ a $(N+L)\times N$ cyclic prefix insertion matrix given by
\begin{equation}
 \Am = \left[
\begin{array}{c}
 \mathbf{0_{L,N-L}}\quad \Id_L \\
%\hline
\Id_{N}
\end{array}
\right].
\end{equation}
The channel matrix representing the link from the MBS to the $j^{th}$ MUE, after the cyclic prefix removal operation, is defined as $\Tc(\hv_{\text{mm}}^{(1,j)}) \in \mathbb{C}^{N\times (N+L)}$, where $\Tc(\cdot)$ is a Toeplitz operator that returns a Toeplitz matrix built from a given vector, i.e., for $\hv = \left[ h(0) \cdots h(L) \right]$:
\begin{eqnarray} \label{eq:toep}
\Tc(\hv) =
\left[ \begin{array}{cccccc}
h(L) & \cdots & h(0) & 0 & \cdots & 0 \\
0 & \ddots &  & \ddots & \ddots & \vdots \\
\vdots & \ddots & \ddots &  & \ddots & 0 \\
0 & \cdots & 0 & h(L) & \cdots & h(0) \\
\end{array} \right].
\end{eqnarray}
\noindent 
Concerning the second tier, the SBSs adopt a block transmission scheme that will be detailed in Sec. \ref{sec:single}. We assume that an SUE is not different from an MUE with respect to the reception chains, being distinguished merely by the association point (MBS or SBS). Therefore, like the MUEs, the SUEs discard the leading $L$ symbols and perform a DFT at the reception. Naturally, no mask filter is needed at $k^{th}$ SUE, given that, in general, no OFDMA-based transmission can be performed in the second tier without generating cross-tier interference towards the MUEs. Let $\Tc(\hv_{\text{sm}}^{(i,j)}) \in \mathbb{C}^{N\times (N+L)}$ be the matrix representing the channel from the $i^{th}$ SBS to the $j^{th}$ MUE, constructed from the $\hv_{\text{sm}}^{(i,j)}$ channel coefficients.  The matrices $\Tc(\hv_{\text{ms}}^{(1,k)})$, $\Tc(\hv_{\text{ss}}^{(i,k)}) \in \mathbb{C}^{N\times (N+L)}$, representing the link from MBS and the $i^{th}$ SBS to the $k^{th}$ SUE respectively, can be similarly constructed. 

Now, let $\yv_{\text{m}}^{(j)}$, $\yv_{\text{s}}^{(k)}$ be the received $N$-sized vector at the $j^{th}$ MUE and $k^{th}$ SUE, respectively, $\sv_{\text{m}}$ be the MBS input vector of size $N$, composed of $M$ individual zero mean, unit norm symbol vectors $\sv_{\text{m}}^{(j)}, ~ j\in[1,M]$, and $\xv_{\text{s}}^{(i)}$ be the transmit vector at the $i^{th}$ SBS, of size $N+L$, detailed later for clarity. Then, if we let $\nv_{\text{m}}^{(j)}$, $\nv_{\text{s}}^{(k)}\sim\Cc\Nc(0, \sigma^2 \Id_{N})$ be two additive white Gaussian noise (AWGN) vectors, the received signals at the $j^{th}$ MUE and $k^{th}$ SUE can be expressed as 
\begin{align}
\begin{split}
\yv_{\text{m}}^{(j)} &= \Bm_{j}\Fm \bigg(\Tc(\hv_{\text{mm}}^{(1,j)})\Am \Fm^{-1} \sv_{\text{m}} + ... \\ &\quad + \sum_{i=1}^{K} \Tc(\hv_{\text{sm}}^{(i,j)}) \xv_{\text{s}}^{(i)} + \nv_{\text{m}}^{(j)}\bigg) 
\end{split}  \label{eq:received_single_user_m} \\
\begin{split}
\yv_{\text{s}}^{(k)} &= \Fm \bigg(\Tc(\hv_{\text{ss}}^{(i,k)}) \xv_{\text{s}}^{(i)} + ... \\
&\quad + \sum_{l\neq i}^{K} \Tc(\hv_{\text{ss}}^{(l,k)}) \xv_{\text{s}}^{(l)} + \Tc(\hv_{\text{ms}}^{(1,k)})\Am \Fm^{-1} \sv_{\text{m}}+ \nv_{\text{s}}^{(k)}\bigg). 
\end{split} \label{eq:received_single_user_s}
\end{align}
Note that, in \eqref{eq:received_single_user_s}, we represented $\yv_{\text{s}}^{(k)}$ by separating the useful signal received from the $i^{th}$ SBS from the co-tier interference component generated by the remaining $K-1$ SBSs, operating in the second tier. For $\yv_{\text{m}}^{(j)}$, the all of the second tier transmitted signal is seen as interference.

To simplify the subsequent analysis, consider an equivalent aggregate model that includes all users in the system. Let us start by looking at the first tier. By summing up all the contributions of the MUEs, orthogonal in the frequency domain, the equivalent channel matrix from the MBS to the MUEs is
\begin{equation}\label{eq:channel_p}
    \Hm_{\text{mm}} = \sum_{j=1}^{M} \Bm_{j}\Fm \Tc(\hv_{\text{mm}}^{(1,j)})\Am \Fm^{-1} \in \mathbb{C}^{N \times N}.
\end{equation}
Let us now define 
\begin{equation} \label{eq:channel_s_simp}
\Hm_{\text{sm}}^{(i,\cdot)} = \sum_{j=1}^{M} \Bm_{j}\Fm \Tc(\hv_{\text{sm}}^{(i,j)}) \in \mathbb{C}^{N \times (N+L)},
\end{equation}
then the equivalent aggregated interference channel from the SBSs to the MUEs is constructed as
\begin{eqnarray} \label{eq:interf_chann}
\Hm_{\text{sm}} = \left[ \begin{array}{cccccc} \Hm_{\text{sm}}^{(1,\cdot)}, \dots, \Hm_{\text{sm}}^{(K,\cdot)}   \end{array} \right] \in \mathbb{C}^{N \times K(N+L)}.
\end{eqnarray}
Switching our focus to the second tier, let $\Hm_{\text{ss}}^{(i,k)} = \Tc(\hv_{\text{ss}}^{(i,k)})$. Then, by defining
\begin{equation} \label{eq:direct_chann_small}
\widetilde{\Hm}_{\text{ss}}=\left[ \begin{array}{cccccc}
\Hm_{\text{ss}}^{(1,1)} & \cdots & \Hm_{\text{ss}}^{(1,K)} \\
\Hm_{\text{ss}}^{(2,1)} & \cdots & \Hm_{\text{ss}}^{(2,K)} \\
\vdots & \ddots & \vdots \\
\Hm_{\text{ss}}^{(K,1)} & \cdots & \Hm_{\text{ss}}^{(K,K)} \\
\end{array} \right] \in \mathbb{C}^{KN \times K(N+L)},
\end{equation}
the equivalent aggregated channel from the SBSs to the SUEs can be written as
\begin{equation}
\Hm_{\text{ss}} = (\Id_K \otimes \Fm)\widetilde{\Hm}_{\text{ss}} \in \mathbb{C}^{KN \times K(N+L)}.
\end{equation}
The interfering link from the MBS to the SUEs is \linebreak obtained by following the same approach. Let \linebreak $\Hm_{\text{ms}}^{(1,k)}= \Tc(\hv_{\text{ms}}^{(1,k)})\Am \Fm^{-1} \in \mathbb{C}^{N \times N}$. By defining
\begin{equation}
\widetilde{\Hm}_{\text{ms}}=\left[ \begin{array}{cccccc}
\Hm_{\text{ms}}^{(1,1)} \\
\Hm_{\text{ms}}^{(1,2)} \\
\vdots \\
\Hm_{\text{ms}}^{(1,K)} \\
\end{array} \right] \in \mathbb{C}^{KN \times N},
\end{equation}
we can write the equivalent aggregated channel as
\begin{equation}
\Hm_{\text{ms}} = (\Id_K \otimes \Fm)\widetilde{\Hm}_{\text{ms}} \in \mathbb{C}^{KN \times N}.
\end{equation}  
Now, we define $\yv_{\text{m}} = \displaystyle \sum_{j=1}^{M} \yv_{\text{m}}^{(j)}$ as the aggregated received vector at the MUEs of size $N$, and $\yv_{\text{s}}\triangleq[\yv_{\text{s}}^{(1)\text{T}},\dots,\yv_{\text{s}}^{(K)\text{T}}]^{\text{T}}$ as the aggregated received vector at the SUEs of size $KN$. We also define $\xv_{\text{s}}\triangleq[\xv_{\text{s}}^{(1)\text{T}},\dots,\xv_{\text{s}}^{(K)\text{T}}]^{\text{T}}$ as the aggregated transmit vector at the SBSs, of size $K(N+L)$. The equivalent signal model is then obtained as 
\begin{eqnarray} 
\yv_{\text{m}}&=& \Hm_{\text{mm}} \sv_{\text{m}} + \Hm_{\text{sm}}\xv_{\text{s}} + \nv_{\text{m}} \label{eq:received_tot} \\
\yv_{\text{s}}& = &\Hm_{\text{ss}}\xv_{\text{s}} + \Hm_{\text{ms}} \sv_{\text{m}} + (\Id_{K} \otimes \Fm)  \nv_{\text{s}} \label{eq:received_tott}.
\end{eqnarray}
Note that, in (\ref{eq:received_tot}) and \eqref{eq:received_tott}, $\displaystyle \nv_{\text{m}} = \sum_{j=1}^{M} \Bm_{j}\Fm \nv_{\text{m}}^{(j)}$ and \linebreak $\nv_{\text{s}}=[\nv_{\text{s}}^{(1)\text{T}},\dots,\nv_{\text{s}}^{(K)\text{T}}]^{\text{T}}$ are the aggregated AWGN vectors of the first and second tier, of size $N$ and $KN$ respectively.

\section{Precoder Design} \label{sec:prec_des}

According to the cognitive overlay paradigm \cite{art:goldsmith2009bs}, the secondary system must protect the primary network from the interference caused by the opportunistic transmission. By looking at \eqref{eq:received_tot}, we see that this implies
\begin{equation} \label{eq:nullcondition}
\Hm_{\text{sm}}\xv_{\text{s}}  = \textbf{0}.
\end{equation}
The transmitted message by the MBS to the MUEs is not known in the secondary system, which disqualifies algorithms like dirty paper coding (DPC) \cite{art:costa83}. Furthermore, the SBSs possess no information about unused resources (time, space or frequency) at the primary system and each MUE is a single antenna device. Therefore, traditional techniques to design an interference-free transmission \cite{art:cadambe2008i}-\cite{art:akoum11} can not be implemented in the considered scenario. 

Let $\sv_{\text{s}}^{(i)}$ be the input symbol vector at the $i^{th}$ SBS, detailed later for clarity. Consequently, let $\sv_{\text{s}}\triangleq[\sv_{\text{s}}^{(1)\text{T}},\dots,\sv_{\text{s}}^{(K)\text{T}}]^{\text{T}}$ be the aggregated SBSs' input symbol vector, such that 
\begin{equation}\label{eq:x}
\xv_{\text{s}}=\Em \sv_{\text{s}}
\end{equation}
becomes its precoded version through a linear precoder $\Em$, whose design is discussed in the following. Then \eqref{eq:nullcondition} can be rewritten as
\begin{equation} \label{eq:nullcondition2}
\Hm_{\text{sm}}\Em = \textbf{0}.
\end{equation}
If we assume that each SBS may independently precode its input vector to cancel the interference towards the MUEs, we can express $\Em$ as the direct sum \cite{book:knapp2006} of $K$ precoders
\begin{equation} \label{eq:overall_pre}
\Em = \bigoplus_{i=1}^{K} \Em_i,
\end{equation}
where $\Em_i$ is the precoder at the $i^{th}$ SBS. It is straightforward to see that when the following holds
\begin{equation} \label{eq:new_orthcon}
\Hm_{\text{sm}}^{(i,\cdot)}\Em_{i} = \textbf{0}, \hspace{4mm} \forall i \in [1,K],
\end{equation}
\eqref{eq:nullcondition2} is always satisfied, if perfect knowledge of $\Hm_{\text{sm}}^{(i,\cdot)}$ is available at the $i^{th}$ SBSs. Thus, the SBSs do not need to share any information related to the cross-tier interference channels towards the MUEs to create $\Em$. This results in a simpler architecture as well as in a lower backhaul signaling. As a consequence, we can focus on the $i^{th}$ SBS to devise $\Em_i$ and then apply \eqref{eq:overall_pre} to find the desired overall precoder.  Moreover, we note that a CSI measurement is valid only throughout the coherence time of the channel of interest, \linebreak e.g., $\Hm_{\text{sm}}^{(i,\cdot)}$. Therefore, we must seek for one-shot strategies that do not require iterative procedures between the SBSs and the SUEs/MUEs to derive the precoding/decoding matrices, such as the IA-based solutions in \cite{art:el-keyi11} and references therein. At this stage, we assume perfect CSIT related to the interfering links from the SBSs towards the MUEs. In the second part of the work, the impact of imperfect CSIT will be analyzed.

\subsection{Single SBS/SUE Precoder Design} \label{sec:single}

\begin{figure}[!h]
\centering
\includegraphics[width=\columnwidth]{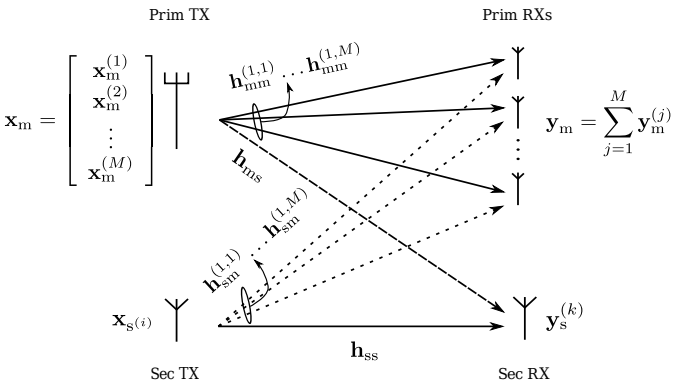}
	\vspace{-5mm}
	\caption{OFDMA downlink interference channel model, single SBS.}
	\label{fig:scenario_small}
\end{figure}

\noindent
We first focus on the pair given by the $i^{th}$ SBS and its SUE $k$, thus a scenario as in Fig. \ref{fig:scenario_small}, i.e., $K=1$. 
In \cite{conf:cardoso2008v}, we proposed a linear precoder called VFDM to solve the interference cancelation problem in a single primary, single secondary scenario. 
This kind of situation meant that the overall channel matrix was of Toeplitz structure $\Tc(\hv)$ with $\hv = [h_1, \dots, h_L]$, allowing a linear precoder based on a Vandermonde matrix \cite{book:golub1996m} to be constructed from the roots of the interfering channel's $\hv$ polynomial given by
\begin{equation}
S(z) = \sum_{i=0}^L h_i z^{L-i}.
\end{equation}
Unfortunately, unlike in \cite{conf:cardoso2008v}, herein the considered SBS deals with a multi-user OFDMA downlink. Due to the multiple interfering links from the SBS to the MUEs, no polynomial representation of the equivalent channel is possible, and the Vandermonde-subspace based result is not directly applicable to our case. Nevertheless, the null-space precoder idea can still be used, as shown in the following.

By looking at (\ref{eq:new_orthcon}), we note that, if this precoder exists, then it must lie within the kernel of ${\Hm}_{\text{sm}}^{(i,\cdot)}$. In the considered scenario, the redundancy introduced at the MBS, to combat the multipath interference, ensures that $\rank{({\Hm}_{\text{sm}}^{(i,\cdot)})} = N$, thus $\dimV{(\kerV{(\Hm_{\sx\mx}^{(i,\cdot)})})}=L$. Therefore, the non-emptiness of the kernel is guaranteed by the block transmission structure adopted in the first tier, i.e., OFDMA, and a solution to \eqref{eq:new_orthcon} can be found. Now, let $\Hm_{\text{sm}}^{(i,\cdot)} = \Lm _{\text{sm}}^{(i,\cdot)}\Qm_{\text{sm}}^{(i,\cdot)}$ be the LQ decomposition of the equivalent channel matrix representing the interfering link between the $i^{th}$ SBS and the MUEs, where $\Lm_{\text{sm}}^{(i,\cdot)} \in \mathbb{C}^{N\times (N+L)}$ is a lower triangular matrix and $\Qm_{\text{sm}}^{(i,\cdot)} \in \mathbb{C}^{(N+L)\times (N+L)}$ is a unitary matrix given by
\begin{equation} \label{eq:precoder_Q}
\Qm_{\text{sm}}^{(i,\cdot)} \triangleq \left[~ \qv_1 ~|~ \qv_2 ~|~ \cdots ~|~ \qv_{N+L} ~\right]. 
\end{equation}
By construction, we know that the last $L$ orthonormal columns of $\Qm_{\text{sm}}^{(i,\cdot)\text{H}}$ lie within $\ker{(\Hm_{\text{sm}}^{(i,\cdot)})}$. Therefore, if we define
\begin{equation} \label{eq:precoder}
\Em_i \triangleq \left[~ \qv_{N+1} ~|~ \cdots ~|~ \qv_{(N+L)-1} ~|~ \qv_{N+L} ~\right] \in \mathbb{C}^{(N+L) \times L}, 
\end{equation}
\noindent		
we have an orthogonal precoder that fulfills (\ref{eq:new_orthcon}). If we substitute \eqref{eq:precoder} into \eqref{eq:overall_pre}, we see that the precoder $\Em$ is obtained as a ${{K(N+L) \times KL}}$ matrix, whose dimension determines the size of the previously defined aggregated zero mean, unit norm SBSs' input symbol vector $\sv_{\text{s}}$, i.e., $KL$. 

We first focus on the macro-cell. If we plug \eqref{eq:x} into (\ref{eq:received_tot}), then we obtain
\begin{equation}\label{eq:finaly_p}
    \yv_{\text{m}} = \Hm_{\text{mm}} \sv_{\text{m}} + \bm{\nu}_{\text{m}}, 
\end{equation}
\noindent
realizing the desired cross-tier interference cancelation. Note that, in \eqref{eq:finaly_p}, $\bm{\nu}_{\text{m}} \in \Cc\Nc(0, \sigma_n^2\Id_N)$ is the DFT of the AWGN vector $\nv_{\text{m}}$, having the same size and statistic. 

Concerning the received signal at the considered SUE, we can rewrite \eqref{eq:received_single_user_s} as
\begin{equation} \label{eq:finaly_s}
    \yv_{\text{s}}^{(k)} = \Fm \Hm_{\text{ss}}^{(i,k)} \Em_i \sv_{\text{s}}^{(i)} + \Fm \Hm_{\text{ms}}^{(1,k)} \sv_{\text{m}} + \bm{\nu}_{\text{s}}^{(k)},
\end{equation}
\noindent
where the co-tier interference component is absent, being the focus of the section on a single SBS/SUE pair. In \eqref{eq:finaly_s}, $\Em_i$ is a linear precoder as defined in \eqref{eq:precoder}, $\Hm_{\text{ms}}^{(1,k)} \sv_{\text{m}}$ is the cross-tier interference coming from the MBS and $\bm{\nu}_{\text{s}}^{(k)} \in \Cc\Nc(0, \sigma_n^2\Id_{N})$ is the DFT of $\nv_{\text{s}}^{(k)}$.

At this stage, the dimension of $\sv_\sx^{(i)}$, zero mean, unit norm input symbol vector at the $i^{th}$ SBS, is clear. In particular, the size of $\sv_\sx$, i.e., $KL$, implies that $\sv_\sx^{(i)}$ is an $L$-sized vector. Consequently, each SBS has an implicit upper bound ($L$) on the number of input symbols that can be precoded by $\Em_i$. This, together with the perfect CSIT assumption, is the cost of the cross-tier interference cancelation constraint induced by the overlay cognitive approach. This guarantees absence of cross-tier interference at the MUEs if perfect CSIT is available at the SBSs, differently from the underlay approach adopted by the cognitive beamforming solutions discussed in Sec. \ref{sec:intro}. We note that, unlike other interference management schemes that exploit the spatial degrees of freedom by the use of multiple antennas, i.e. zero forcing (ZF), and/or special decoding strategies at the receiver, i.e., IA, the proposed technique requires only one antenna per SBS and MUE and legacy OFDM decoding at the latter. In fact, the interference towards the primary system is canceled by adopting a precoder $\Em_i$ that opportunistically exploits the redundancy introduced by the MBS to combat inter symbol (ISI) and inter block interference (IBI), e.g. the cyclic prefix. In the following, we start from these findings to analyze the multi SBS/SUE scenario described in Sec. \ref{sec:model}.

Finally, we note that, the complexity of the LQ decomposition of an $N\times(N+L)$ matrix, e.g ${\Hm}_{\text{sm}}^{(i,\cdot)}$, is \linebreak $\mathcal{O}(N(N+L)^2 - (N+L)N^2+ (N+L)^3)$ \cite{book:golub1996m}. Consequently, \linebreak a centralized approach to find the null space of \linebreak the aggregated cross-tier interference channel matrix, i.e., $\Hm_{\text{sm}}$ in \eqref{eq:interf_chann}, would require an LQ decomposition of complexity $\mathcal{O}(N[K(N+L)]^2 - [K(N+L)]N^2+ [K(N+L)]^3)$, growing exponentially with $K$. Therefore, the distributed nature of the proposed $\Em_i$ precoding not only reduces the backhaul signaling requirements, but dramatically decreases the complexity of the processing in the second tier, where $K$ low complexity LQ decompositions are performed in parallel to derive the $K$ individual precoders.

\subsection{Multi SBS/SUE VFDM Precoder Design} \label{sec:multi}

As seen in Sec. \ref{sec:single}, the SBSs separately design the precoders $\Em_i$, $\forall i \in [1,K]$, such that the overall precoder $\Em$ as shown in (\ref{eq:overall_pre}) successfully satisfies (\ref{eq:nullcondition}). As a consequence, we can rewrite the signal model in (\ref{eq:received_tot}) and \eqref{eq:received_tott} as
\begin{eqnarray} 
\yv_{\text{m}}&=& \Hm_{\text{mm}} \sv_{\text{m}} + \nv_{\text{m}} \label{eq:received_tot_21}\\ 
\yv_{\text{s}}&=& \Hm_{\text{ss}}  \Em  \sv_{\text{s}} + \Hm_{\text{ms}} \sv_{\text{m}} + \bm{\nu}_{\text{s}}.,\label{eq:received_tot_22} 
\end{eqnarray}
with $\bm{\nu}_{\text{s}}=[\bm{\nu}_\sx^{(1)\text{T}},\dots,\bm{\nu}_\sx^{(1)\text{T}}]^{\text{T}}$. We focus on the second tier and, for clarity, we simplify the notation by introducing
\begin{equation} \label{eq:equivalent}
\overline{\Hm}_{\text{ss}} = \Hm_{\text{ss}}  \Em \in \mathbb{C}^{KN \times KL}.
\end{equation}
\noindent
The structure of the received signal is the same for any SUE, hence we can rewrite (\ref{eq:finaly_s}) for the multi-user case as
\begin{equation} \label{eq:finaly_3}
    \yv_{\text{s}}^{(k)} = \Fm \Hm_{\text{ss}}^{(i,k)} \sv_{\text{s}}^{(i)} + \overline{\Hm}_{\text{ss}}^{([i], k)} \sv_{\text{s}}^{[i]} + \Fm \Hm_{\text{ms}}^{(1,k)} \sv_{\text{m}} + \bm{\nu}_{\text{s}}^{(k)},
\end{equation}
in which we identify a useful component, two interfering terms and the thermal noise. In (\ref{eq:finaly_3}), $\overline{\Hm}_{\text{ss}}^{([i], k)} \sv_{\text{s}}^{[i]} \in \mathbb{C}^{N \times (K-1)L}$ represents the co-tier interference experienced by each SUE. Clearly, the performance of the second tier hinges on the mutual interference between the SBSs and is strongly interference limited as $K$ increases. Note that, as in the single user case, the absence of cooperation between the two tiers implies that the MBS' interference on the SUEs is always present. Consequently, in this scenario, each SUE deals with a stronger interference if compared to the single SBS case in Sec. \ref{sec:single}. To address this issue we assume that the SBSs may communicate over an infinite-capacity backhaul realizing a coordinated network MIMO system. Despite being hardly realistic, this assumption is usually made in similar scenarios for first studies on newly-proposed algorithms, to put focus on ultimate bounds of such solutions and achieve a better understanding of their potential \cite{art:gesbert10, art:karayakali06}. The cooperating SBSs can be therefore modeled as a MIMO broadcast channel (MIMO-BC), whose capacity is given by DPC \cite{art:caire03}, a difficult to implement technique. Because of its complexity, many suboptimal but linear strategies have been introduced lately. Accordingly, we propose to address the co-tier interference problem at the cooperating SBSs by adding one linear suboptimal precoding layer, resulting in an overall cascaded \linebreak MU-VFDM precoder, as detailed in the following sections.

\subsection{Dimensionality Problem and Linear Techniques} \label{sec:lin}
Having solved the cross-tier interference problem, now we devote our attention to mitigating the co-tier interference by means of a linear suboptimal precoder. As such, in (\ref{eq:received_tot_22}), we focus on the SBSs' transmission by isolating the term  $\overline{\Hm}_{\text{ss}}$ of dimension $KN \times KL$, as defined in (\ref{eq:equivalent}). Note that, in any block transmission system, the added redundancy $L$ to the block of $N$ useful symbols is always such that $\frac{L}{N} < 1$, for matters of efficiency. As seen in Sec. \ref{sec:single}, \linebreak MU-VFDM imposes a dimensionality constraint to the transmitters in the second-tier since each SBS precodes up to $L$ input symbols while each SUE receives $N$ symbols. This implies that a direct application of techniques such as zero forcing beamforming (ZFBF) \cite{book:gold05} or block diagonalization (BD) \cite{art:spen04} is not possible, since both require that the number of columns (transmit dimensions) of the channel matrix be bigger or equal than the number of rows (receive dimensions). Regularized inverse beamforming (RIBF) \cite{art:peel2005} is applicable, but it achieves poor performance at high SNR, due to the aforementioned dimensionality issue ($N>L$ received symbols) that yields a very poor condition number to the equivalent channel representation built upon Toeplitz matrices. Matched filter (MF) precoding \cite{conf:demig2008} performs similarly, being largely suboptimal at high SNR. It is known from \cite{art:visw2002}, and for the multiple beams case from \cite{art:sharif2005}, that opportunistic random beamforming (ORBF) based techniques are able to yield the optimal capacity scaling of $KL\log\log KN$ in dense networks with a large number of receivers. Unfortunately, in our scenario the ratio $\frac{N}{L}$ is such that we can not achieve good performance using these techniques. In general, most of the results in the literature regarding linear precoding techniques under given optimization criteria assume only one antenna/symbol at the receiver. For this reason, a direct extension of these techniques is not possible. 

Algorithms like iterative regularized block diagonalization (IRBD) \cite{art:stan2008} deal with multiple symbols/antennas at each receiver. The higher experienced diversity gain is due to the suppression of the interference only between the symbols received by two different receivers. These algorithms perform better than other techniques that rely on the single antenna/symbol assumption. On the other hand, they require a joint receiver decoding, with a consequent increase in the complexity of the \linebreak receivers' architecture. 

Simpler solutions, implemented to deal with an arbitrary number of dimensions at each receiver, are user/antenna selection based algorithms. It is known that by scheduling only a subset of antennas or eigenmodes \cite{art:spen04} to be served using a classical ZFBF, the achievable sum-rate is asymptotically optimal \cite{art:yoo2006}. In spite of this, the condition for the asymptotic optimality is never met in a MU-VFDM system, thus neither an exhaustive search of the optimal subset nor a faster and suboptimal greedy selection algorithm \cite{art:dimic05} can achieve good results. 

Looking at the schemes presented thus far, we note that the inherent dimensionality constraint limits the performance of the second tier, in terms of both achievable sum-rate and complexity of the SBSs/SUEs. Starting from this consideration, we propose a low complexity solution to overcome the dimensionality constraint and manage multi-user interference in the following section. 

\subsection{RIBF Flexible Network Solution} \label{sec:zfbf}
Consider a flexible approach to the second-tier deployment in which the network designer can  modify the dimensionality of the system by installing more antennas at each SBS/SUE, or alternatively by increasing the SBS' density. We let $\gamma_{tx}, \gamma_{rx}$ be two parameters such that $\gamma_{tx}L, \gamma_{rx}N \in \mathbb{N}$ are the number of transmit and receive dimensions respectively, with $L$ and $N$ fixed due to the OFDMA symbol structure. This way, the network designer can tune $\gamma_{tx}$ and $\gamma_{rx}$ to capitalize on the flexibility of the model, effectively changing the number of available channels for the transmission, and obtaining different operating scenarios. For instance, when $\gamma_{tx}=1$ and $\gamma_{rx}$ grows large, the system experiences a large increase of the number of receive dimensions, i.e., implying a greater number of SUEs (or SUEs' antennas) from which the best ones to serve are selected, and this represents the condition under which ORBF is optimal (a very "tall" overall channel matrix). Conversely, if $\gamma_{rx}$ is kept constant ($\gamma_{rx}=1$ for simplicity) and we let $\gamma_{tx}$ increase, the SBSs can exploit the abundance of transmit dimensions to achieve a higher transmit diversity, thanks to the greater number of considered channels. Another interesting configuration is given by $\gamma_{tx}=N$ and $\gamma_{rx}=L$, that is a network where the number of transmit and receive dimensions coincides, i.e., channel inversion based techniques such as ZFBF and RIBF become efficient in terms of degrees of freedom exploitation. These strategies do not require iterative or greedy algorithms to be implemented, and thus, represent an attractive solution to manage the co-tier interference by means of a one-shot technique. In particular, it is known from \cite{art:peel2005} that RIBF offers better performance for a wider class of channels, regularizing the matrix to be inverted whenever its condition number is poor. Consequently, in the following we will focus on RIBF, and we note that it can be implemented effectively in the considered scenario if the dimensionality constraint is overcome, thus if the following holds
\begin{equation} \label{eq:zfbf_cond}
\gamma_{tx} L \geq \gamma_{rx} N.
\end{equation}
\noindent
Then, without loss of generality, we let $\gamma_{rx}=1$ and $\gamma_{tx}$ increase. In particular, we note that this preserves the legacy number of antennas per SUE, i.e., 1, and their disjoint decoding strategy. Due to the large number of SBSs (or antennas per SBS), we consider a uniform power allocation strategy to reduce the computational burden for the SBSs. We remark that, thanks to the $\gamma_{rx}$ and $\gamma_{tx}$ tuning, the second tier is characterized by a greater number of channels. As a consequence, in the new setup,  $\sv_{\text{s}}$ is a vector of size $\gamma_{tx}KL$, $\Em \in \mathbb{C}^{\gamma_{tx}K(N+L) \times \gamma_{tx}KL}$ and $\overline{\Hm}_{\text{ss}} \in \mathbb{C}^{KN \times \gamma_{tx}KL}$.  At this stage, we can define
\begin{equation} \label{eq:zf_precoder}
\Phim=\overline{\Hm}_{\text{ss}}^{\text{H}}(\frac{\sigma_n^2}{P_s} \Id_{KN} + \overline{\Hm}_{\text{ss}} \overline{\Hm}_{\text{ss}}^{\text{H}})^{-1}
\end{equation}
as the joint RIBF precoder, with $\Phim \in \mathbb{C}^{\gamma_{tx}KL \times KN}$. Then, if we let $\uv_{\text{s}} \in \mathbb{C}^{KN \times 1}$ be a new aggregated SBSs' input symbol vector, such that $\sv_{\text{s}} = \Phim \uv_{\text{s}}$ we can rewrite the signal model given by \eqref{eq:received_tot_21} and \eqref{eq:received_tot_22} as
\begin{eqnarray} 
\yv_{\text{m}} &=& \Hm_{\text{mm}} \sv_{\text{m}} + \nv_{\text{m}} \label{eq:received_tot_31}\\ 
\yv_{\text{s}} &=& \Hm_{\text{ss}}  \Wm \uv_{\text{s}} + \Hm_{\text{ms}} \xv_{\text{m}} + \bm{\nu}_{\text{s}}, \label{eq:received_tot_32}
\end{eqnarray}
where 
\begin{equation}
\Wm=\frac{\Em \Phim}{\sqrt{\trace(\Em \Phim \Phim\herm \Em \herm)}} \in \mathbb{C}^{\gamma_{tx}K(N+L) \times KN}
\end{equation}
is the overall normalized MU-VFDM cascaded precoder, such that $\trace{(\Wm\herm\Wm)}=1$. We emphasize that, the cascaded precoder structure is intrinsically different from that of our previous work in \cite{conf:cardoso2008v}, even for the $K,M=1$ case. In fact, the use of an outer linear precoding scheme, while preserving the interference cancelation condition towards the first tier, substantially changes the dimensionality of the system.

\section{Numerical Analysis} \label{sec:numerical}

In this section, we present a numerical performance analysis of the proposed technique. Please note that, according to \linebreak Sec. \ref{sec:model} and \ref{sec:prec_des}, the matrices $\Hm_{\text{sm}}$ and $\Em$ are not composed of i.i.d. random entries, but are strongly structured. No closed form of the eigenvalue/eigenvector distribution is available, and a purely theoretical performance analysis can not be carried out. Consequently, we proceed by means of \linebreak Monte Carlo based simulations of the considered \linebreak downlink scenario, comprised of an OFDMA/LTE MBS in the macro-cell with $M=4$ MUEs, and an MU-VFDM based small-cell system. For simplicity, we consider the least resource-demanding extended mode proposed by the standard \cite{rpt:3gpp25.814}, and characterized by $N=128$ subcarriers, a cyclic prefix of length $L=32$, for a total bandwidth of 1.92~MHz. Noise and channel vectors are generated as described in Sec. \ref{sec:model}. First we assume that perfect CSI is available at the SBSs, afterwards we admit for the presence of noisy channel estimations yielding imperfect CSIT. Note that, if not stated otherwise in the text, we do not consider any interference from the MBS to the SUEs to isolate the effect of the MU-VFDM precoder on second tier's performance. In particular, this assumption is crucial to evaluate the effect of the imperfect channel estimation at the SBSs on the performance of the cascaded precoder designed in Sec. \ref{sec:prec_des}. Finally, for the sake of compactness of the notation in our plots, we introduce the \textit{load rate} $\beta$ as the ratio between the number of transmit and receive dimensions as defined in \linebreak Sec. \ref{sec:zfbf}, and given by
\begin{equation} \label{eq:beta}
\beta = \frac{\gamma_{tx}L}{\gamma_{rx}N}. 
\end{equation}

\subsection{Multi-User VFDM}

\begin{figure}[!h]
	\centering
\includegraphics[width=.92\columnwidth]{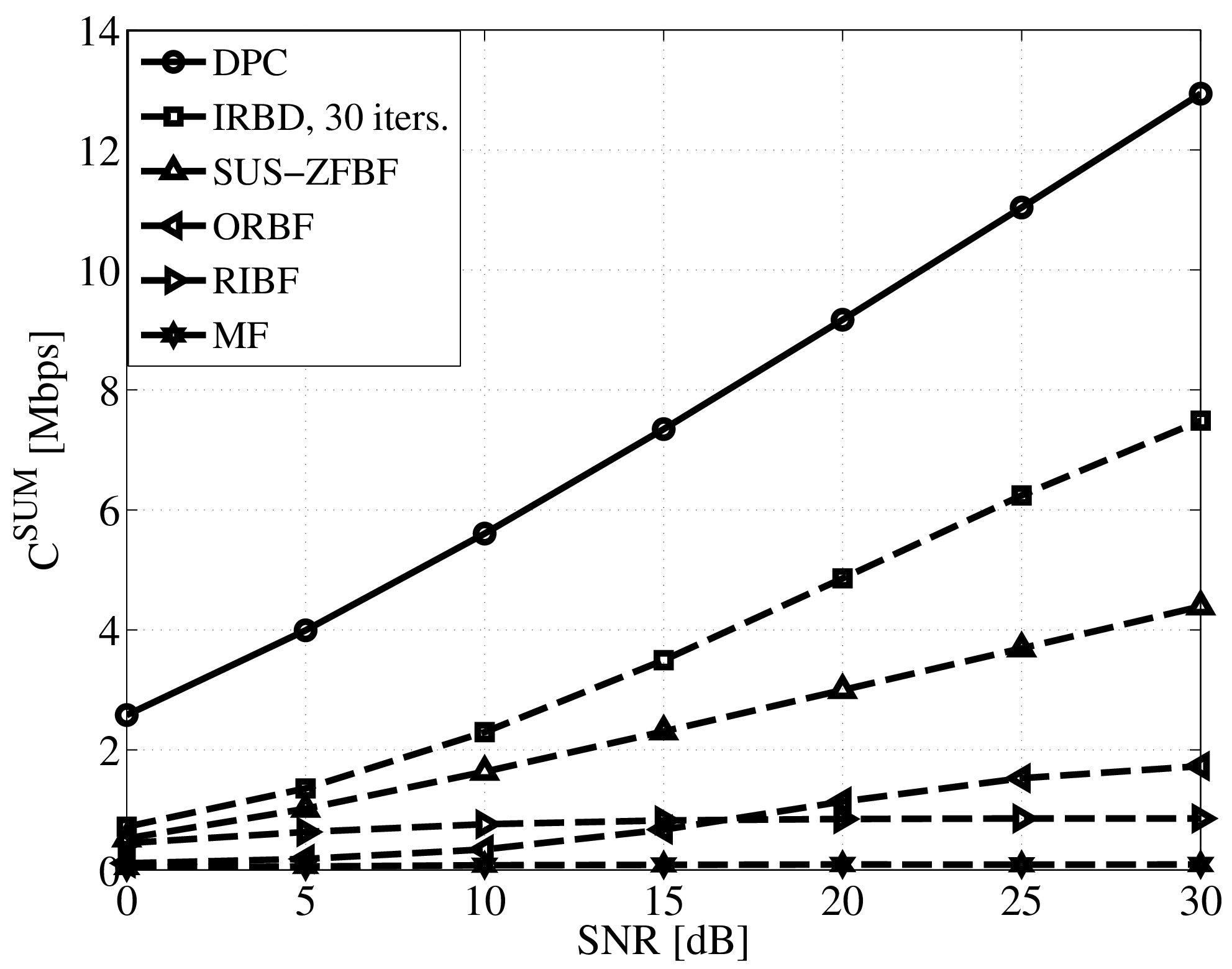}
	\vspace{-3mm}
	\caption{Rate of the SBSs for different transmit schemes, $K=3$ ($N=~128, L=32$ and bandwidth of 1.92~Mhz).}
	\label{fig:sec_prec}
\end{figure}

Consider a small-cell system composed by $K=3$ \linebreak SBSs/SUEs. Let us assume that the SBSs null the interference towards the MUEs by (\ref{eq:precoder}), then the upper bound capacity $C^{\text{SUM}}_{\text{DPC}}$ achieved by adopting DPC, for a uniform power allocation is as follows \cite{book:tse2005f}
\begin{equation} \label{eq:dpc_rate}
%C^{\text{SUM}}_{\text{DPC}} = \frac{\Bc}{N+L}\mathbb{E}\left[\log_{2} \left| \Id_{KN} + \text{SNR}(\Id_{KN}+ \frac{P_m}{\sigma_n^2} \Hm_{\text{ms}}\Hm_{\text{ms}}\herm)^{-1}\overline{\Hm}_{\text{ss}}\overline{\Hm}_{\text{ss}}^{H}\right| \right].
C^{\text{SUM}}_{\text{DPC}} = \frac{\Bc}{N+L}\mathbb{E}\left[\log_{2} \left| \Id_{KN} + \left(\frac{N+L}{\sigma^2 L\gamma_{tx}}\right)P_s\overline{\Hm}_{\text{ss}}\overline{\Hm}_{\text{ss}}^{H}\right| \right],
\end{equation}
where $\Bc$ is the considered bandwidth and $P_m$ and $P_s = \frac{P_m}{K}$ are the power per transmit symbol at the MBS and at each SBS, respectively. Note that, the adopted model implies that the total transmit power per tier is the same, i.e., $P_m(N+L)$, and the larger $K$ becomes, the lower the power budget available at each SBS. This is imposed to model the second tier in compliance with the lower energy consumption requirements that the SBSs will likely have w.r.t. a legacy MBS in 4G networks \cite{art:hoydis11}. In Fig. \ref{fig:sec_prec}, $C^{\text{SUM}}_{\text{DPC}}$ is compared to the achievable ergodic sum-rate $C^{\text{SUM}}$ of MU-VFDM where the $\Phim$ stage precoding is obtained by some of the linear precoding strategies presented in \linebreak Sec. \ref{sec:lin}, for SNR~$\in [0,30]$, including the semi-orthogonal user selection ZFBF (SUS-ZFBF) algorithm proposed in \cite{art:yoo2006}. 
The behavior of the considered linear precoding schemes shows a big rate offset when compared to the upper bound given by $C^{\text{SUM}}_{\text{DPC}}$, and this confirms what has been discussed in Sec. \ref{sec:lin}.

\begin{figure}[!h]
	\centering
	\includegraphics[width=.92\columnwidth]{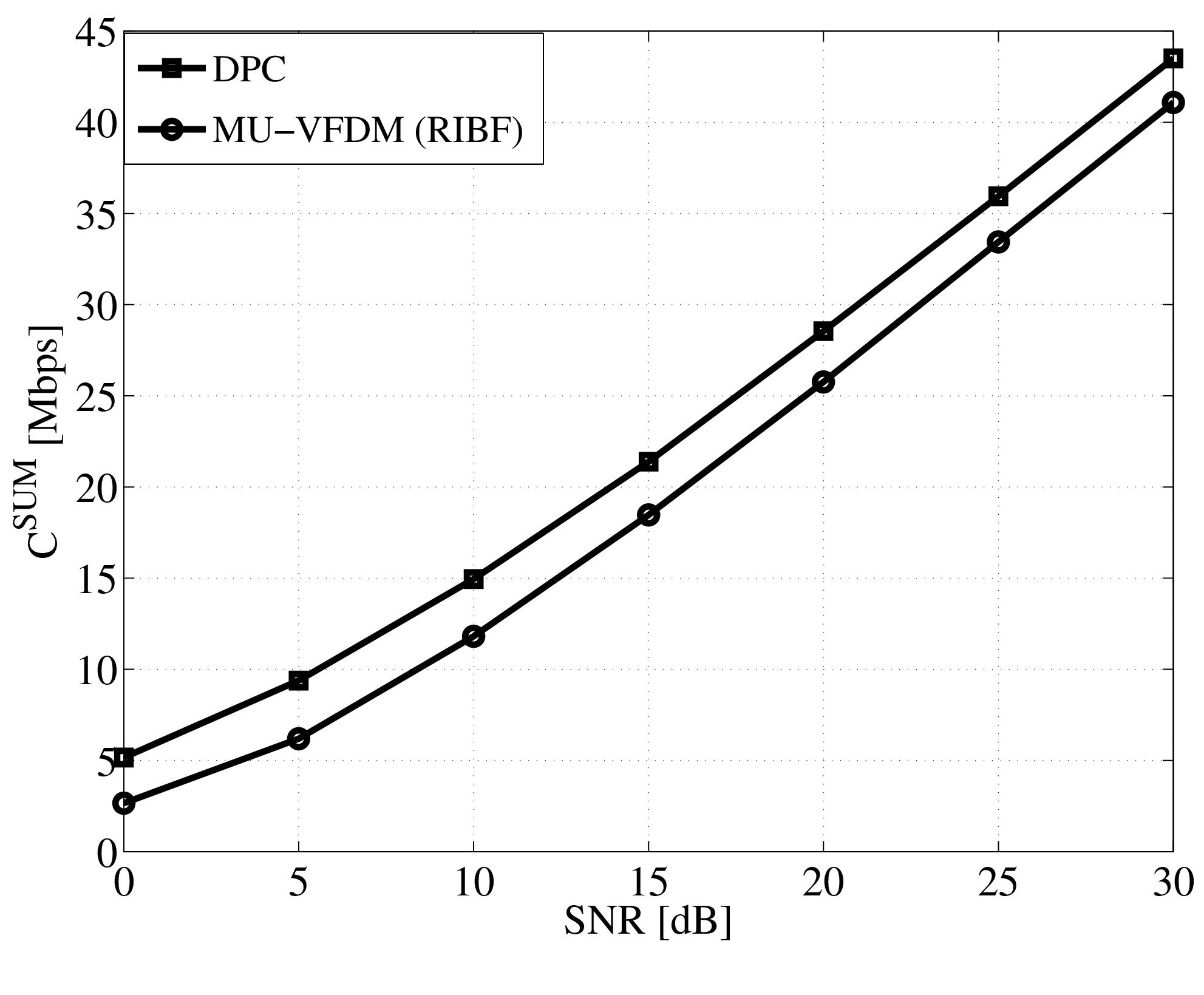}
	\vspace{-5mm}
	\caption{Achievable rate of the SBSs with the MU-VFDM (RIBF) flexible network solution, $K=3$, $\beta=2.5$ ($N=128, L=32$ and bandwidth of 1.92~Mhz).}
	\label{fig:sec_zfbf}
\end{figure}

To compute the sum-rate of the small-cell system implementing MU-VFDM with the RIBF outer precoder, $C^{\text{SUM}}_{\text{RIBF}}$, we need to evaluate the SINR for each of the $KN$ received symbols at the SUEs. Let $\Phim = [\phiv^{(1)},\dots,\phiv^{(KN)}]$. Let $\overline{\hv}_{\text{ss}}^{(j)}=$~$[[\overline{\Hm}_{\text{ss}}]_{j1},\dots, [\overline{\Hm}_{\text{ss}}]_{j\gamma_{tx}KL}]$ denote the $j^{th}$ row of $\overline{\Hm}_{\text{ss}}$, then we can write
\begin{equation} \label{eq:sinr}
\text{SINR}_{(\text{s}), j} = \frac{|\overline{\hv}_{\text{ss}}^{(j)}\phiv^{(j)}|^2}{\sum_{i \neq j}^{KN} |\overline{\hv}_{\text{ss}}^{(j)}\phiv^{(i)}|^2 + \frac{\trace(\Wm \Wm^{\text{H}})\sigma_n^2}{P_s K(N+L)}}, \quad \forall j \in [1,KN]
\end{equation}
where the dimension of $\overline{\Hm}_{\text{ss}}$ depends strictly on the value assumed by $\beta$. Then, it is straightforward to see that for a $K$-SBS system the achievable sum-rate, when perfect CSIT is available, is given by
\begin{equation}
C^{\text{SUM}}_{\text{RIBF}} = \frac{\Bc}{N+L} \sum_{j=1}^{KN}\log_2(1 + \text{SINR}_{(\text{s}),j}).
\end{equation}
In Fig. \ref{fig:sec_zfbf} we illustrate a comparison between $C^{\text{SUM}}_{\text{RIBF}}$ and $C^{\text{SUM}}_{\text{DPC}}$, for a load rate of $\beta = 3$, confirming that the proposed technique has comparable performance to \linebreak state-of-the-art solutions. 
In particular, due to the inherent simplicity and flexibility of the proposed solution, the SBSs' performance can be made arbitrarily close to the upper bound, by increasing the number of dimensions at the transmit side. We remark that, the complexity of the linear precoding techniques outperforming RIBF in Fig. \ref{fig:sec_prec} prevents their implementability for $\beta > \frac{L}{N}$. This consideration further motivates the proposed solution for the multi-user VFDM dense network deployment. 

\subsection{Imperfect CSIT}

In Sec. \ref{sec:prec_des}, we showed that when perfect CSIT is available at the SBSs, an interference nulling precoder $\Em$ can be designed. However, in a realistic implementation, each transmitter in the system performs noisy channel estimations, yielding imperfect CSIT. Therefore, in this section, we seek for a deeper understanding of the impact of the CSIT acquisition on the overall network performance. We recall that in Sec. \ref{sec:multi} we assumed an infinite-capacity backhaul. This allows us to target our efforts on the analysis of the effect of a noisy channel estimation onto the performance of the \linebreak two-tiered network. The study of the achievable performance for a two-tiered network operating under limited backhaul capacity, and the impact of the quantization of the CSI, will be a subject of future work. The design of a suitable channel estimation procedure is out of the scope of this work as well, thus, for simplicity, we assume a classic training/transmission scheme as in \cite{art:hass03}.

\begin{figure}[!h]
	\centering
	\includegraphics[width=.95\columnwidth]{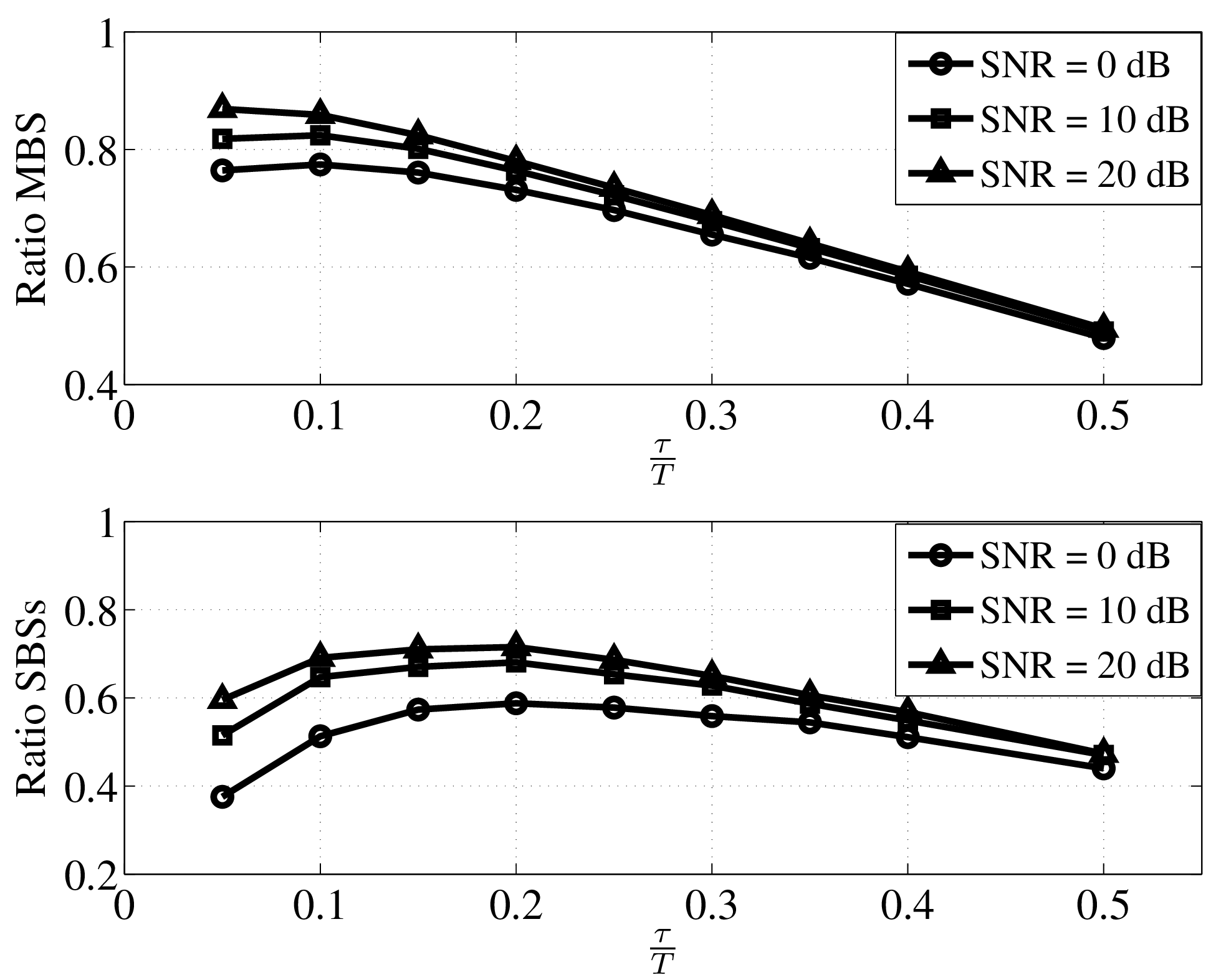}
	\vspace{-3mm}
	\caption{Ratio between the rate obtained with imperfect CSIT and the rate
obtained with perfect CSIT for MBS and SBSs as the SNR changes, $\beta=1$ and $K=3$ ($N=64, L=16$ and bandwidth of 0.96~Mhz).}
	\label{fig:imperf_macro_small}
\end{figure}

Consider a block fading channel model where a channel estimation is valid throughout the duration of the coherence time $T$. The channel estimations are performed during a period $\tau \leq T$, hence the available time for transmission is upper bounded by $T - \tau$. During the training phase the devices broadcast orthonormal sequences of known pilot symbols of equal power. Each channel observation can be expressed as
\begin{equation}
\rv = \sqrt{\rho \tau}\hv + \nv,
\end{equation}
where $\hv$ is the channel vector, $\rho$ is the transmit power and $\nv \sim \Cc\Nc(0, \sigma_n^2\Id_{(L+1)})$ models the effect of the Gaussian noise at the devices' circuitry, driving the experienced SNR at the estimating device. Each device computes the minimum mean-square error (MMSE) estimate of $\hv$, by evaluating the observation $\rv$. At this stage, $\hv$ can be expressed by means of two components, i.e., an estimate $\hat{\hv}$ and an independent Gaussian error $\tilde{\hv}$ \cite{art:hoydis10}, that is
\begin{equation}
\hv = \hat{\hv} + \tilde{\hv}. 
\end{equation}
Without the perfect CSI assumption at the SBSs, the zero interference constraint in \eqref{eq:nullcondition} can no longer be satisfied, thus the SBSs may generate interference towards the MUEs. If we denote the $j^{th}$ row of $\Hm_{\text{mm}}$ as $\hv_{\text{mm}}^{(j)}=$~$[[\Hm_{\text{mm}}]_{j1},\dots, [\Hm_{\text{mm}}]_{jN}]$, and the $j^{th}$ row of $\Hm_{\text{sm}}$ as $\hv_{\text{sm}}^{(j)}=$~$[\Hm_{\text{sm}}]_{j1},\dots, [\Hm_{\text{sm}}]_{j\gamma_{tx}KL}]$, then the SINR per received symbol at the MUEs reads
\begin{equation}
\text{SINR}_{(\text{m}), j} = \frac{P_m K|{\hv^{(j)}_\text{mm}}|^2}{\sum_{i=1}^{KN} |\hv_{\text{sm}}^{(j)}\phiv^{(i)}|^2 + \sigma_n^2}, \forall j \in [1,N].
\end{equation}
Note that, the imperfect CSI at the SBSs has an impact on the general design of $\Wm$, worsening the SINR per received symbol at the SUEs, due to channel estimation effects and increased co-tier interference component. Therefore, \eqref{eq:sinr} does not hold for this case and each SUE experiences an effective SINR value \cite{art:hass03} per received symbol given by
\begin{equation}
\text{SINR}_{(\text{s}), j}=\frac{\bigg(\frac{|\overline{\hv}_{\text{ss}}^{(j)}\phiv^{(j)}|^2}{\sum_{i \neq j}^{KN} |\overline{\hv}_{\text{ss}}^{(j)}\phiv^{(i)}|^2 + \frac{\trace(\Wm \Wm^{\text{H}})\sigma_n^2}{P_s K(N+L)}} \bigg)^2 \tau}{1 + (1 + \tau)\frac{|\overline{\hv}_{\text{ss}}^{(j)}\phiv^{(j)}|^2}{\sum_{i \neq j}^{KN} |\overline{\hv}_{\text{ss}}^{(j)}\phiv^{(i)}|^2 + \frac{\trace(\Wm \Wm^{\text{H}})\sigma_n^2}{P_s K(N+L)}}},
\end{equation}
$\forall j \in [1,KN]$, where we assume that the same transmit power is used for training and data symbols. Then, the sum-rate of the MBS and SBSs respectively is
\begin{eqnarray}
C^{\text{SUM, I}}_{\text{m}} &=& \frac{T-\tau}{T(N+L)}\sum_{j=1}^N \log_2(1 + \text{SINR}_{(\text{m}), j}) \\
C^{\text{SUM, I}}_{\text{s}} &=& \frac{T -\tau}{T(N+L)} \sum_{j=1}^{KN} \log_2(1 + \text{SINR}_{(\text{s}), j}). 
\end{eqnarray}
To reduce Monte Carlo simulation times, we consider $N=64$ active subcarriers, cyclic prefix length of $L=16$ and a load rate of $\beta=1$. In Fig. \ref{fig:imperf_macro_small}, the ratio between the sum-rate obtained with imperfect CSIT and the sum-rate obtained with perfect CSIT is computed for the MBS and the SBSs as different $\tau/T$ proportions are chosen, for SNR $\in \{0, 10, 20\}$~dB. 
Consider the MBS. We note that the optimal $\tau$ hinges on the SNR and, in particular, $\tau=0.1T$ is the optimal value in the low to medium SNR regime. On the other hand, the result for SNR$=20$~dB shows that the pre-log factor dominates the sum-rate in this regime, and the best performance is obtained for the minimum value considered in the simulation, i.e., $\tau=0.05T$. Interestingly, the rate loss experienced by the MBS for SNR$=0$~dB is around $22\%$. Thus, the \linebreak cross-tier interference cancelation provided by MU-VFDM shows a promising robustness to imperfect CSI even if the experienced SNR is very low. Switching our focus to the second tier, we see that the impact of the channel estimation errors at the SBSs on the effectiveness of the \linebreak co-tier interference mitigation is larger. As a result, the SBSs experience a non-negligible sum-rate loss for imperfect CSIT, especially at very low SNR. However, differently from what we have seen for the MBS, the optimal value  for $\tau$ does not show a clear dependence on the SNR, being consistently $\tau=0.2T$ throughout the considered SNR range. In particular, we note how the sum-rate loss varies slowly with $\tau$. This implies that small variations on the available time for the channel estimation w.r.t. the optimal $\tau$ are acceptable by the SBSs, allowing for faster suboptimal channel estimations if necessary.
  
\begin{figure}[!h]
	\centering
	\includegraphics[width=.95\columnwidth]{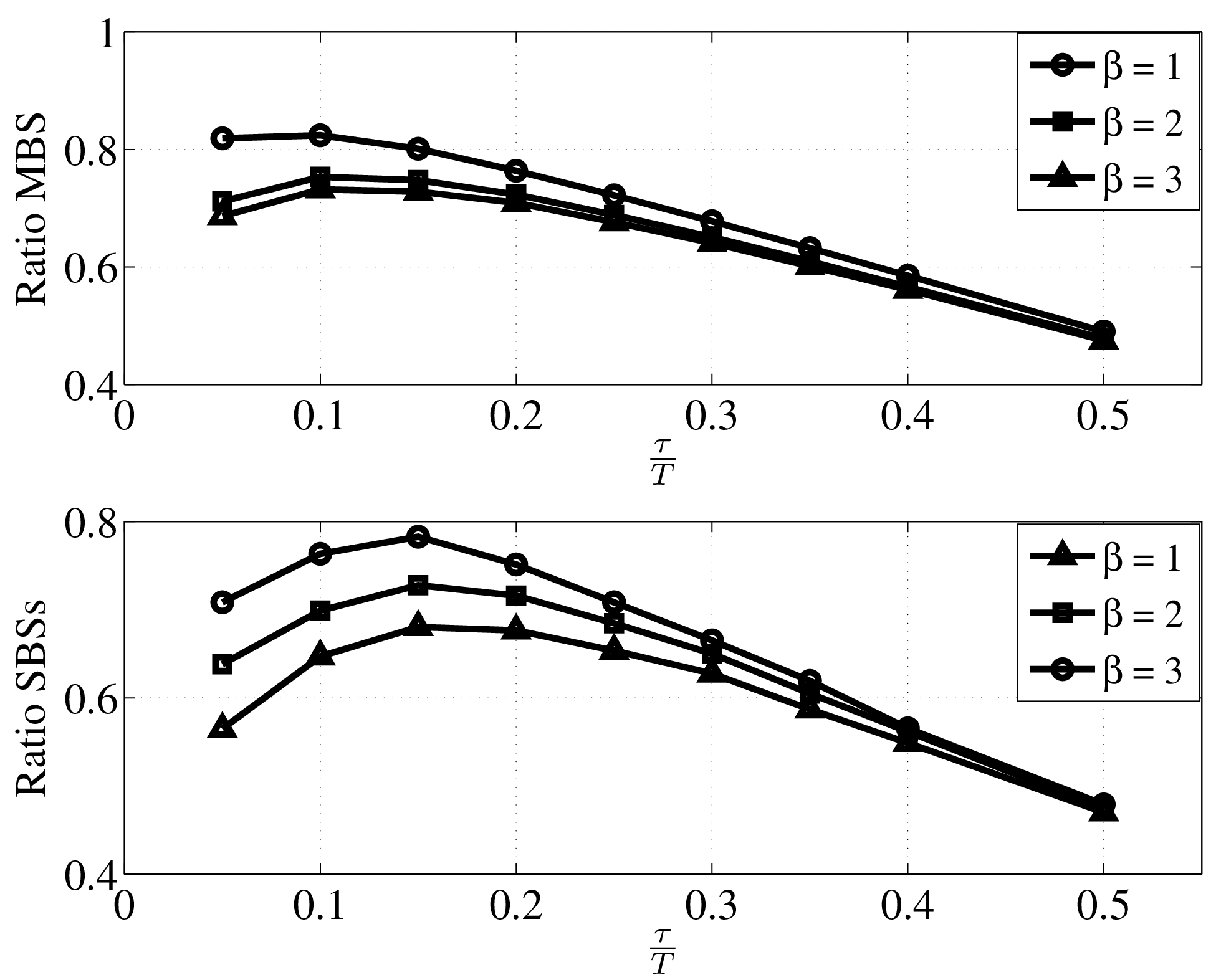}
	\vspace{-3mm}
	\caption{Ratio between the rate obtained with imperfect CSIT and the rate
obtained with perfect CSIT for MBS and SBSs as $\beta$ changes, SNR $=10$ dB and $K=3$ ($N=64, L=16$ and bandwidth of 0.96~Mhz).}
	\label{fig:beta_loss_sub}
\end{figure}

\begin{figure}[!h]
	\centering
	\includegraphics[width=.95\columnwidth]{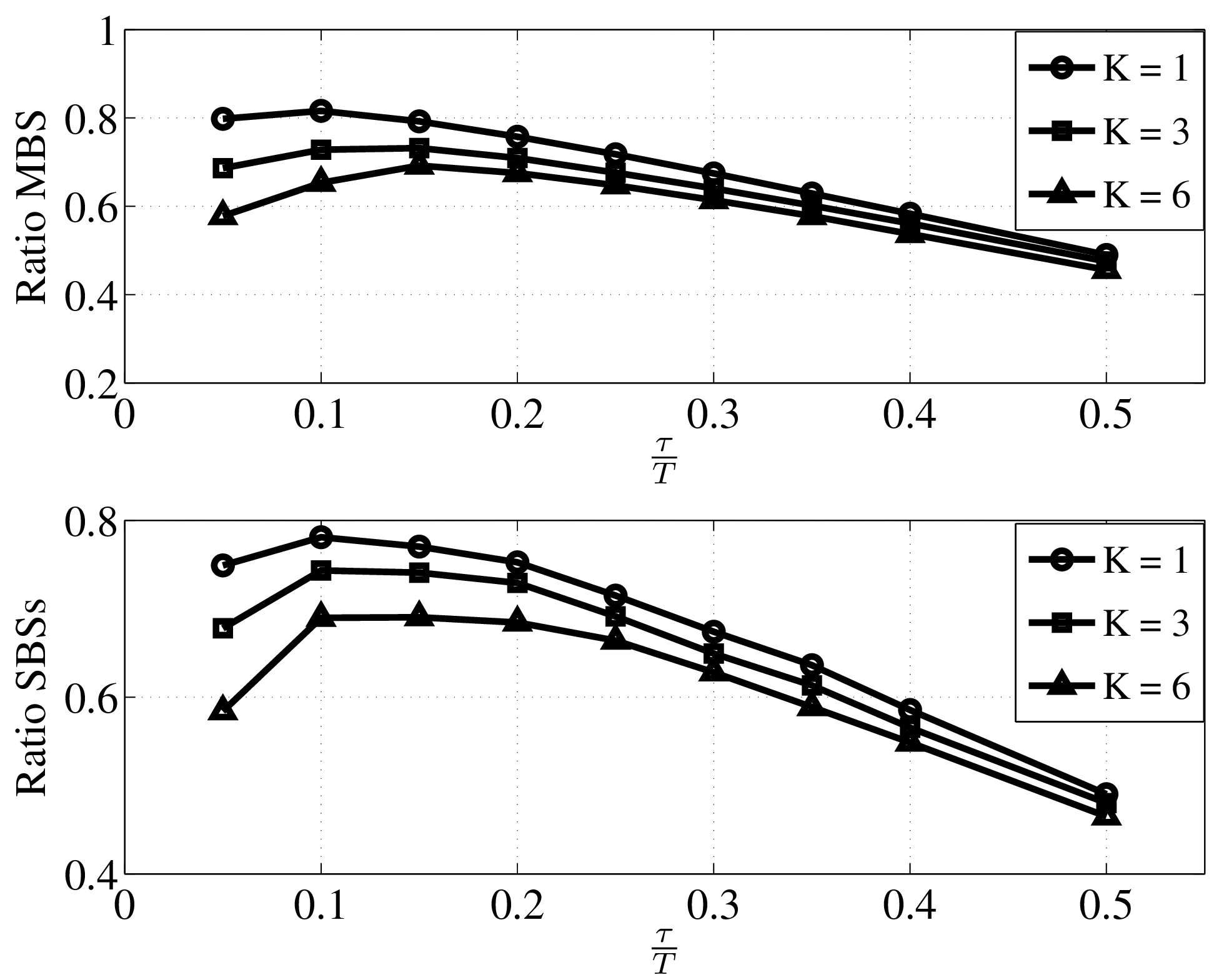}
	\vspace{-3mm}
	\caption{Ratio between the rate obtained with imperfect CSIT and the rate
obtained with perfect CSIT for MBS and SBSs as $K$ changes, SNR~$=10$~dB and $\beta=1$ ($N=64, L=16$ and bandwidth of 0.96~Mhz).}
	\label{fig:K_loss_sub}
\end{figure}

To conclude the analysis on the impact of the imperfect CSIT on the performance of the two-tiered network, we test how MU-VFDM performs as the number of transmit dimensions in the second tier increases. There are two choices at hand: either we modify the ratio between the number of transmit and receive dimensions, i.e., $\gamma_{rx}$ and $K$ fixed and $\beta$ increases, or simply deploy more SBS/SUE pairs, i.e., $\gamma_{rx}$ and $\beta$ fixed and $K$ increases. Therefore, we let the load rate $ \beta \in \{1, 2, 3\}$ in Fig. \ref{fig:beta_loss_sub} (with $\gamma_{rx}=1$, $K=3$), and the number of SBSs $K \in \{1,3,6\}$ in Fig. \ref{fig:K_loss_sub} (with $\gamma_{rx}=1$, $\beta=3$). We assume a constant SNR$=10$~dB. 
\vspace{-2mm}
We first focus on the MBS sum-rate loss. By comparing the two cases, MU-VFDM confirms its robustness to imperfect CSIT and effectiveness for what concerns the cross-tier interference cancelation, regardless of the adopted approach. In particular, we note that $\tau=0.1T$ is optimal for every tested configuration. The sum-rate loss of the SBSs shows a similar trend for the two considered approaches, despite the difference in the optimal value for $\tau$, i.e., $\tau=0.1T$ in Fig. \ref{fig:beta_loss_sub} and $\tau=0.15T$ in Fig. \ref{fig:K_loss_sub}. Nevertheless, we notice that the sum-rate loss for the SBSs increases as $K$ increases, but remarkably shows a decreasing behavior as $\beta$ increases. This interesting result is due to the higher transmit diversity gain experienced by the SBSs as $\beta$ increases. If the number of transmit dimensions is largely greater than the number of receive dimensions, the diversity gain can compensate the rate loss due to the reduced co-tier interference mitigation provided by MU-VFDM in the second-tier for imperfect CSIT, showing the potential of a densely deployed second-tier adopting the proposed technique.

\subsection{Comparison with existing solutions}
\begin{figure}[!h]
	\centering
	\includegraphics[width=\columnwidth]{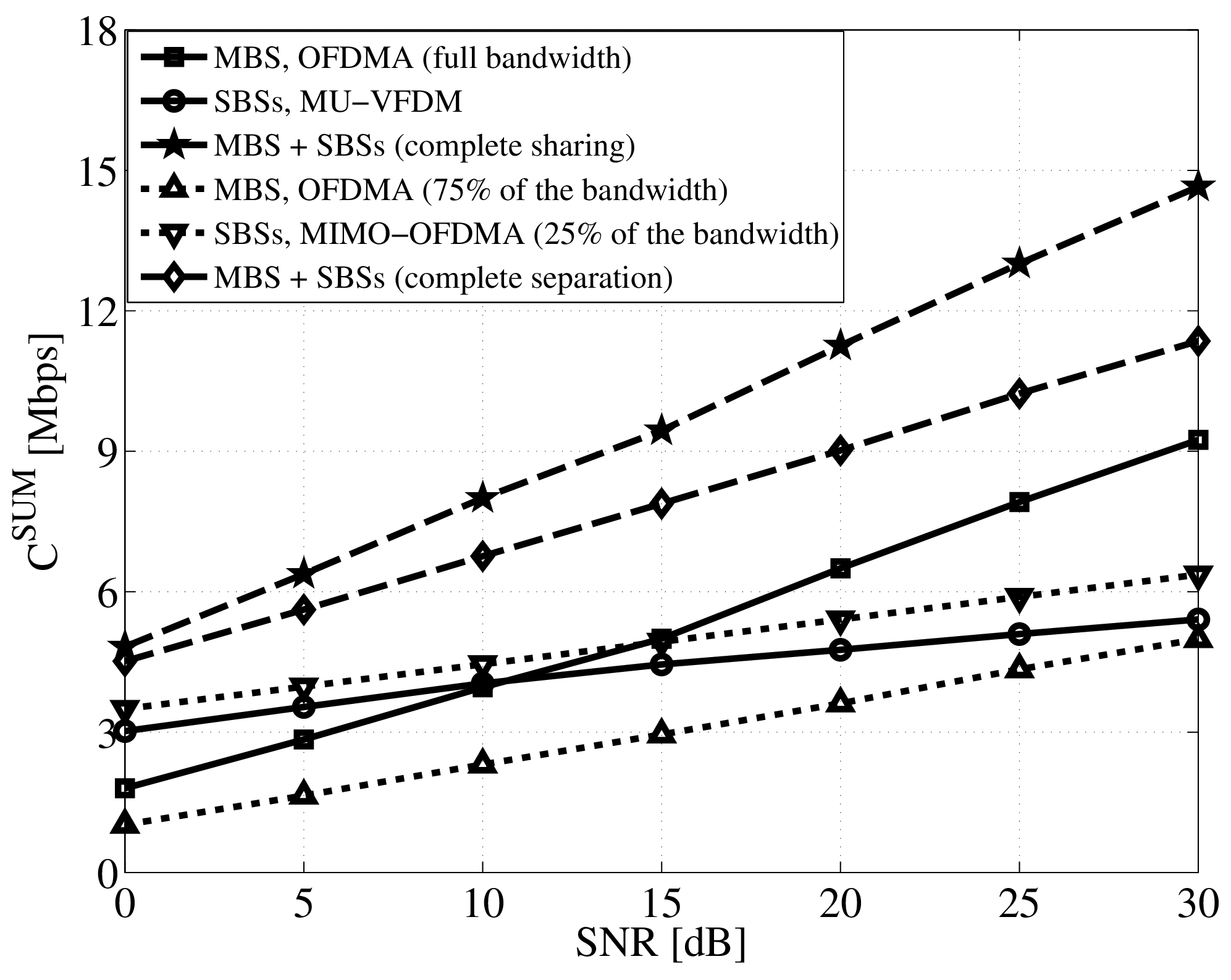}
	\vspace{-3mm}
	\caption{Achievable rate of a two-tiered network, $K=6$, $\beta=3$ and cross-tier interference MBS $\rightarrow$ SUEs ($N=64, L=16$ and bandwidth of 0.96~Mhz). Perfect CSIT.}
	\label{fig:perf_orth}
\end{figure}

\begin{figure}[!h]
	\centering
	\includegraphics[width=\columnwidth]{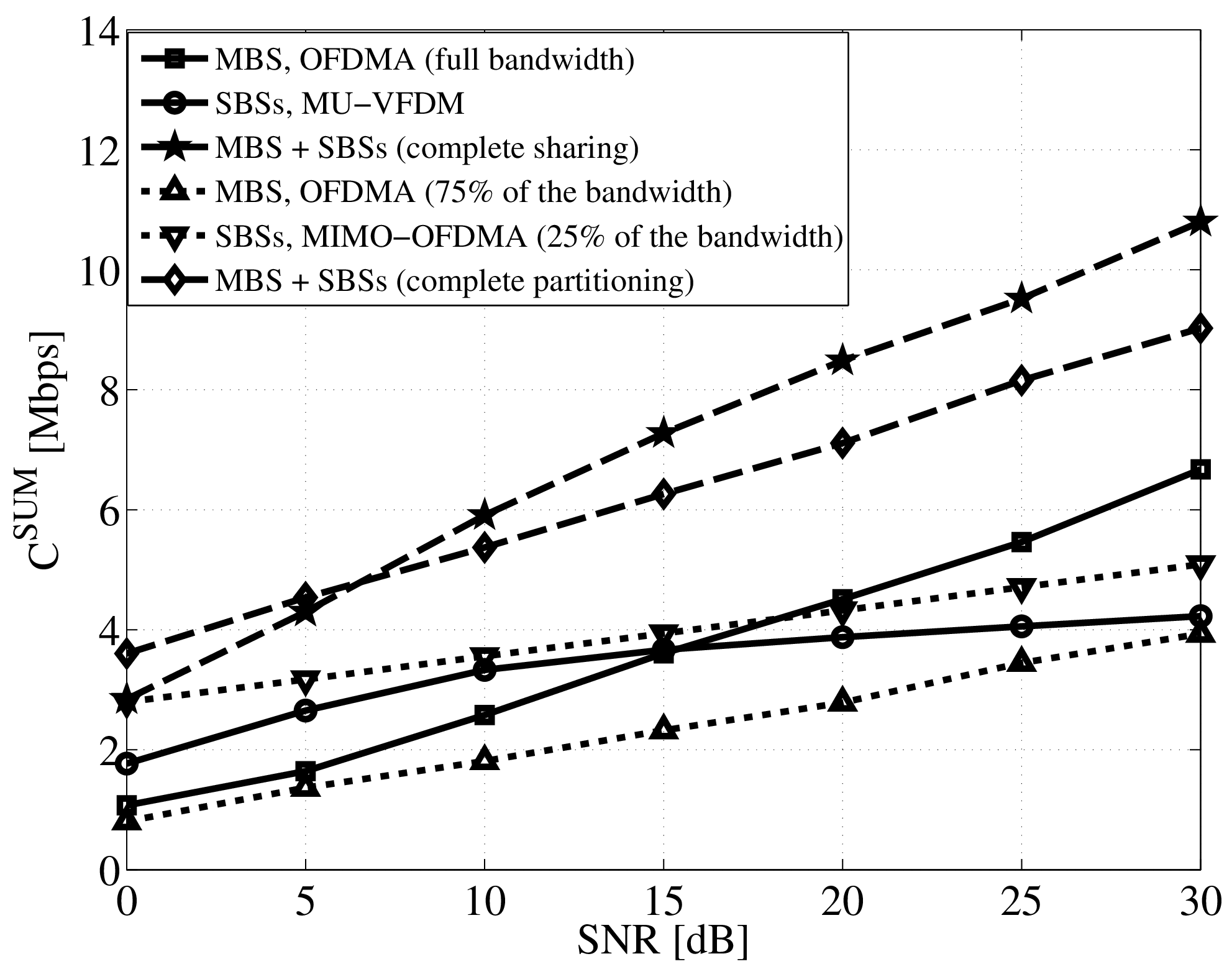}
		\vspace{-3mm}
	\caption{Achievable rate of a two-tiered network, $K=6$, $\beta=3$ and cross-tier interference MBS $\rightarrow$ SUEs ($N=64, L=16$ and bandwidth of 0.96~Mhz). Imperfect CSIT.}
	\label{fig:imperf_orth}
\end{figure}

In this section, we evaluate the performance of the proposed scheme by comparing MU-VFDM to state-of-the-art approaches that allow the deployment of a two-tiered network~\cite{conf:andrews10}: 1) \textit{complete sharing}, 2) \textit{partial sharing}, 3) \textit{complete separation}. MU-VFDM allows the coexistence of SBSs and MBS inside the same area, canceling the interference from the former to the latter, adopting a complete sharing approach. Among the aforementioned bandwidth management schemes, only the complete separation approach guarantees zero interference from the SBSs to the MUEs. Therefore, for a fair comparison, we focus on this approach and divide the available bandwidth in two portions assigned exclusively to the MBS and the SBSs. Considering the values introduced previously, i.e., $N=64$ and $L=16$, this implies that both the \linebreak MU-VFDM and the complete separation based system transmit over a bandwidth $\Bc=0.96$ MHz. As seen in Sec. \ref{sec:single}, by implementing MU-VFDM, each SBS can transmit up to $L$ input symbols from each SBS' antenna. On the other hand, the MBS transmits $N$ input symbols, i.e., the number of considered subcarriers. Consequently, in the complete separation approach, we assign a bandwidth $\Bc_s=\frac{\Bc L}{N}$ to the SBSs and $\Bc_m = \Bc - \Bc_s$ to the MBS. By means of this division, we ensure that each SBS' antenna is transmitting the same number of symbols as in MU-VFDM. Moreover, in order to exploit all the available transmit dimensions, we assume that the SBSs perform a network MIMO-OFDMA transmission towards the SUEs, adopting a ZF precoding such that no linear processing at the SUEs is required, as in MU-VFDM. Note that, a legacy OFDMA transmission is performed by the MBS as described previously. As a last remark, differently from what we have assumed so far, we assume that in MU-VFDM the SUEs suffer from full interference from the MBS. This allows for a more realistic and fair comparison, accounting both for advantages and drawbacks of the two different bandwidth management approaches. We let $\beta=3$ and $K=6$. In Fig. \ref{fig:perf_orth}, the achievable rate of the two schemes for perfect CSIT is presented. 
The complete sharing approach implemented through MU-VFDM shows a clear advantage over the complete separation scheme, at it provides a larger overall sum-rate of the two-tiered network at all SNR regimes. This is achieved despite the large impact of the cross-tier interference from the MBS to the SUEs, clearly noticeable at medium and high SNR values. This remarkable result motivates a further comparison, when only imperfect CSIT is available. In Fig. \ref{fig:imperf_orth}, we see that MU-VFDM  achieves a slightly worse performance if compared to the previous case, even if the overall sum-rate of the two-tiered network is still higher than the performance of complete separation scheme, for SNR values greater than $7$~dB. 
Due to the nature of the cascaded precoder a wrong channel estimation deteriorates the performance especially for low SNR values. On the other hand, the advantage for other SNR regimes is evident and promising, despite the high impact that, as before, the MBS' cross-tier interference has on the performance of the second-tier for medium and high SNR. Therefore, MU-VFDM is able to exploit efficiently the higher multiplexing gain provided by the complete sharing approach, at the expense of a slightly worse performance for low SNR if compared to the complete separation strategy, for the imperfect CSIT case. Consequently, the coexistence in the two-tiered network can be achieved, effectively enhancing the spectral efficiency and the capacity per area, for both CSIT assumptions.

\section{Conclusion} \label{sec:conclusion}

In this work, we presented a DSA cognitive overlay technique, called MU-VFDM, that allows the deployment of SBSs inside the coverage area of a preexisting MBS. We focused on the coexistence over the same bandwidth between a downlink LTE MBS and an SBS system, to illustrate both the feasibility and the performance of MU-VFDM. The network MIMO assumption made a potentially interference limited system become a \linebreak MIMO-BC. Thanks to this fact, several linear precoding techniques involving cooperation between transmitters have been taken into account, and the inherent dimensionality constraint due to the structure of the precoder $\Em$ has been identified. The search for a suitable scheme brought us to the proposed flexible RIBF based approach presented in Sec. \ref{sec:zfbf}. Increasing the number of transmit dimensions, while keeping the receiver layout, is a viable way to design a system that overcomes the dimensionality problem and achieves relevant performance in terms of sum-rate. Such a system design can be realized either by extra antenna installation, denser SBS deployment or a flexible combination of both. The relaxation of the perfect CSIT assumption at the SBSs results in rate loss experienced by both systems, due to the imperfectly devised precoder. The best compromise between training and data symbols has been investigated, for various SNR values, as well as the best performing strategy to deploy a dense network for the imperfect CSIT case. Finally, a comparison with \linebreak state-of-the-art techniques has shown a consistent advantage of the proposed technique for a large range of SNR values, both for perfect and imperfect CSIT case. The results presented herein reinforce our previous findings and confirm that \linebreak MU-VFDM can be used to deploy SBSs and MBS coexisting inside the same coverage area, sharing the same band.

The analysis of the performance of this scheme under limited backhaul capacity assumption is matter of our future research, along with the impact of a partial cooperation between the SBSs. Moreover, we will move from a fully coordinated to a clustered network MIMO scenario, to find different and more practically implementable ways to manage the \linebreak co-tier interference while guaranteeing the cross-tier interference cancelation.
\bibliographystyle{unsrt}
\bibliography{centralized_solution}

\vspace*{-2\baselineskip}

\begin{IEEEbiography}[{\includegraphics[width=1in,clip, keepaspectratio]{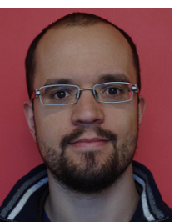}}]
{Marco Maso} received the bachelor’s degree in 2005 and the M.Sc. degree in Telecommunications Engineering in 2008, both from University of Padova, Italy, and is currently pursuing his Ph.D degree at University of Padova and Sup\'elec. He worked on projects dealing with practical implementations of OFDM packet synchronization  in 2005/06, and DVB-T2 system simulation in 2008/09. He is currently involved in the HENIAC project, studying new techniques for high speed coherent optical communications. His research interests include heterogeneous networks, wireless communications, cognitive radio and embedded devices.
\end{IEEEbiography}

\vspace*{-2\baselineskip}

\begin{IEEEbiography}[{\includegraphics[width=1in,clip, keepaspectratio]{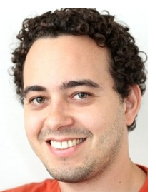}}]
{Leonardo S. Cardoso} received his electrical engineering and M.Sc. degrees from the Universidade Federal do Cear\'a (UFC), Brazil in 2003 and 2006, and his Ph.D. degree in 2011 from Sup\'elec, France, on Cognitive Radio and Dynamic Spectrum Access. From 2001 to 2006, he worked in several projects for Ericsson Research at the Wireless Telecommunications Research Group, GTEL, Brazil. In 2006, he joined the Eur\'ecom Institute, France, working in projects on heterogeneous networks and real-time MIMO channel performance assessment, significantly contributing to the EMOS MIMO platform. He is currently a research fellow at INRIA, France, developing a large scale Cognitive Radio testbed.
\end{IEEEbiography}

\vspace*{-2\baselineskip}

\begin{IEEEbiography}[{\includegraphics[width=1in,clip, keepaspectratio]{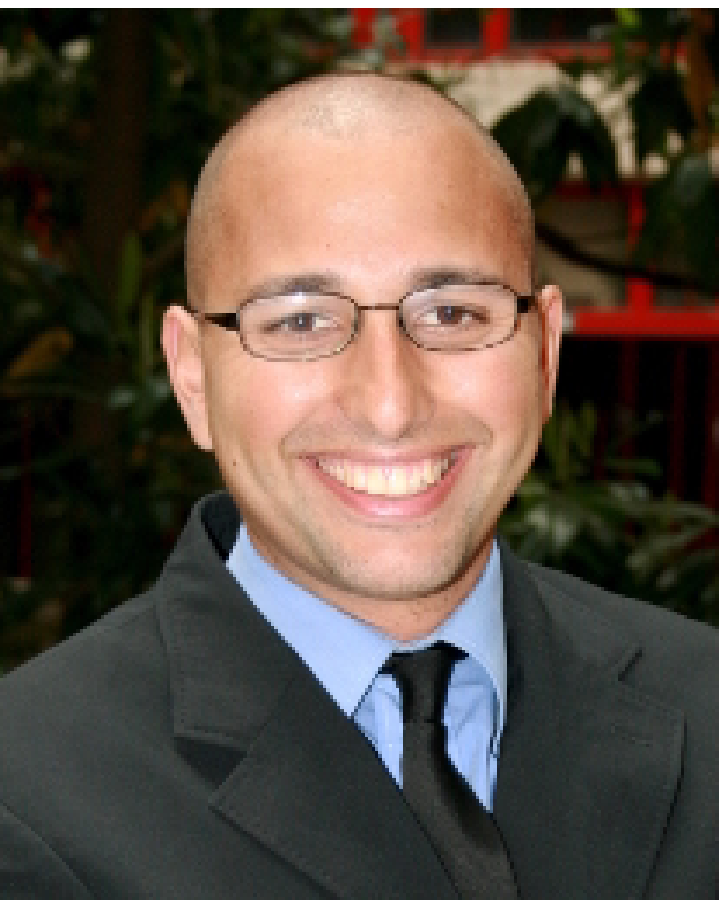}}]
{M\'erouane Debbah} entered the Ecole Normale Sup\'erieure de Cachan (France) in 1996 where he received his M.Sc and Ph.D. degrees respectively. He worked for Motorola Labs (Saclay, France) from 1999-2002 and the Vienna Research Center for Telecommunications (Vienna, Austria) from \linebreak 2002-2003. He then joined the Mobile Communications department of the Institut Eurecom (Sophia Antipolis, France) as an Assistant Professor. Since 2007, he is a Full Professor at Sup\'elec (Gif-sur-Yvette, France), holder of the \linebreak Alcatel-Lucent Chair on Flexible Radio. His research interests are in information theory, signal processing and wireless communications. He is an Associate Editor for IEEE Transactions on Signal Processing. M\'erouane Debbah is the recipient of the "Mario Boella" prize award in 2005, the 2007 General Symposium IEEE GLOBECOM best paper award, the Wi-Opt 2009 best paper award, the 2010 Newcom++ best paper award as well as the Valuetools 2007, Valuetools 2008 and CrownCom2009 best student paper awards. He is a WWRF fellow. In 2011, he received the IEEE/SEE Glavieux Prize Award.
\end{IEEEbiography}

\vspace*{-2\baselineskip}

\begin{IEEEbiography}[{\includegraphics[width=1in,clip, keepaspectratio]{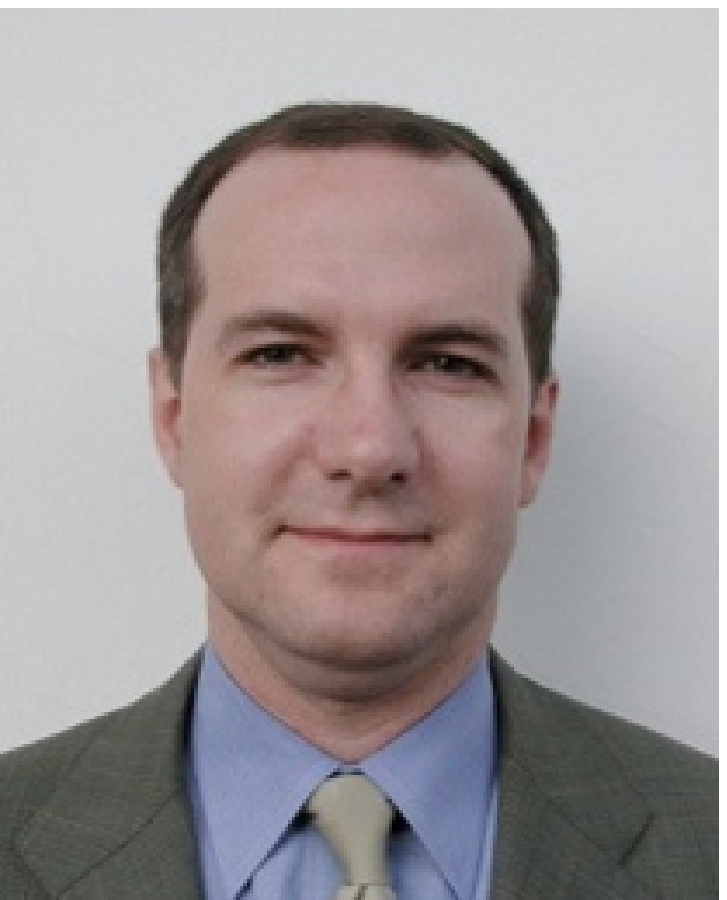}}]
{Lorenzo Vangelista} (SM’02) received the Laurea degree from University of Padova, Padova, Italy, in 1992, and the Ph.D. degree in Electrical and Telecommunication Engineering from University of Padova,in 1995. He subsequently joined the Transmission and Optical Technology Department of CSELT, Torino, Italy. From December 1996 to January 2002, he was with Telit Mobile Terminals, Sgonico (TS), Italy and then, up to May 2003, with Microcell A/S, Copenaghen, Denmark. Until July 2006 he has been with the worldwide organization of Infineon Technologies, as program manager. Since October 2006 he is an Associate Professor of Telecommunication within the Department of Information Engineering of Padova University, Italy. His research interests include signal theory, multi-carrier modulation techniques, cellular networks, wireless sensors and actuators networks and smart-grids.
\end{IEEEbiography}

\end{document}